\documentclass[12pt]{article}

\begin{document}


\newcommand{\TeV}{\,{\rm TeV}}
\newcommand{\GeV}{\,{\rm GeV}}
\newcommand{\MeV}{\,{\rm MeV}}
\newcommand{\keV}{\,{\rm keV}}
\newcommand{\eV}{\,{\rm eV}}
\def\ap{\approx}
\def\bea{\begin{eqnarray}}
\def\eea{\end{eqnarray}}
\def\bi{\begin{itemize}}
\def\ei{\end{itemize}}
\def\be{\begin{enumerate}}
\def\ee{\end{enumerate}}
\def\ler{\lesssim}
\def\gtr{\gtrsim}
\def\beq{\begin{equation}}
\def\eeq{\end{equation}}
\def\haf{\frac{1}{2}}
\def\nn{\nonumber}
\def\p{\prime}
\def\ccg{\cal G}
\def\L{\cal L}
\def\O{\cal O}
\def\R{\cal R}
\def\U{\cal U}
\def\V{\cal V}
\def\W{\cal W}
\def\e{\varepsilon}
\def\slash#1{#1\!\!\!\!\!/}

\setcounter{page}{1}

\title{\Large \bf Softness of Scherk-Schwarz Supersymmetry Breaking}


\author{Hyung Do Kim\\
\\School of Physics\\ Korea Institute for Advanced Study
\\
{\tt hdkim@kias.re.kr} }

\date{\today}


\maketitle

\section*{Abstract}

We investigate how the mass correction appears in 5-D with
Scherk-Schwarz compactification and clarify whether the KK
regularization is reliable method or not. In the extremely sharp
cutoff limit of the 5-D regulator which preserves 5-D Lorentz
invariance, we prove that the one loop correction to the mass does
not depend on the ultraviolet physics for the Scherk-Schwarz
breaking of supersymmetry. This is a unique property of
Scherk-Schwarz breaking which is given by the boundary condition.

\newpage

\section{Introduction}

Theories with extra dimension have been extensively studied by
string theorists and phenomenologists. Extra dimension comes out
naturally from string theory, and it looks desirable to see what
we can get from extra dimensions. The best situation would be to
derive all the physical predictions from an ultimate theory, but
we are far away from the final dream. As a second best way, more
practical and phenomenological bottom up approach has been done
recently. Hierarchy problem, doublet-triplet splitting problem in
GUT, and other problems are reconsidered within the framework of
the models with extra dimensions. Among these, especially in some
models \cite{ADPQ,DPQ}, Scherk-Schwarz (SS) breaking \cite{SS}
plays an important role. Therefore it is very necessary to
understand the physical properties of SS mechanism.

Calculations in the compact extra dimension models involve towers
of Kaluza-Klein (KK) modes and the correct procedure of treating
these fields has not been established. The classical decomposition
of higher dimensional fields into their infinitely many discrete
modes yields several interesting puzzles when we inquire quantum
effects. The well known problem is the incapability of direct
application of dimensional regularization (DR) to it. DR is the
most popular regularization scheme that is widely used due to its
powerful convenience in the calculation. However, the
non-decoupling theorem does not hold for DR, and this makes it
impossible to apply DR to KK towers directly. Matching and
integrating out are essential ingredients in DR, and these
concepts prevent us from direct application of DR to extra
dimensional theory with KK modes.

The choice of the regularization is a matter of taste as long as
the regularization keeps all the symmetries of the theory. In most
cases, the regularization dependence appears only at the next loop
order and is not important as long as the parameters are related
to the measured quantities by the renormalization condition. The
situation is different when we have a theory that can predict
something. If one physical parameter does not have any tree level
bare term for it and only can be induced by loop correction, we
can predict it by measuring other parameters involved in the loop
correction.

Recently one 5-D orbifold model \cite{BHN} with SS supersymmetry
breaking has been suggested, and in the model Higgs mass is
predicted by measuring top quark mass and Z boson mass. One extra
dimension is compactified to $S^1/Z_2\times {Z'}_2$ and boundary
condition is assigned to break supersymmetry such that only the
Standard Model particles are remained as zero modes at tree level.
Yukawa couplings are allowed in the superpotential at two fixed
points, and they generate one loop Higgs mass combined with mass
splitting of top and stop induced by SS supersymmetry
breaking.(There are other Yukawa corrections and gauge coupling
corrections but top-stop gives the dominant contribution.) This
triggered an interesting debate on the so called ``Kaluza-Klein
regularization'' scheme.

In the papers \cite{BHN,AHNSW,HN} they used KK regularization that
sums up all KK modes first and then integrated over 4-D  momentum.
There is no scale above which supersymmetry is manifest, but the
one loop correction to the Higgs mass depends only on the
compactification radius. Position-momentum space mixed formalism
has been used in \cite{AHNSW} to show the exponentially decaying
physics of higher 4-D momentum contributions for spacially
separated propagation in the extra dimension. In \cite{Nilles},
for generic momentum cutoff $\Lambda$ and the truncation of KK
modes at some cutoff $N/R$, it was shown that quadratic
divergences reappear in the calculation. One interesting
observation was that quadratic divergences cancel separately in
the bosonic and the fermionic sectors for KK regularization, i.e.,
for infinite sums of KK modes. In \cite{Kim} it was stressed that
the momentum cutoff and the truncation scale should be the same
due to the 5-D Lorentz invariance, and KK regularization is an
extremely anisotropic regularization that does not keep the 5-D
Lorentz invariance. Furthermore, it was anticipated that the one
loop mass correction to the Higgs would be highly sensitive to the
UV physics if we use the physical isotropic regularization
\cite{Kim}.

There appeared many papers \cite{CPRS,DGJQ,CP,Nomura,Weiner} which
support the finiteness (or UV insensitiveness) of one loop
correction to the Higgs mass originally obtained by weird KK
regularization. In \cite{DGJQ}, Gaussian distribution of fixed
point brane has been used in order to soften the contributions of
heavy KK modes, but the result is obtained only after infinite
sums over KK modes. In \cite{CP}, the calculation has been done in
4-D PV and 4-D DR, and also it needs infinite sums over KK modes
which looks not sensible physically. All these calculations are
based on their own regularizations which may contain hidden
subtraction {\footnote{In 4-D DR with minimal subtraction(MS)
gives a finite answer for scalar mass loop correction and the
divergence is not seen clearly. In odd dimensions, it is more
puzzling since Gamma function $\Gamma (\frac{2-D}{2})$ appearing
in the expression does not have a pole for odd D and the
correction looks finite. In this case, the subtraction of
(cubic/linear($D=5/D=3$)) divergence is hidden in the process of
replacing the divergent integral to finite Gamma function. All
these divergences are seen clearly in the momentum cutoff
regularization.}} and can not rule out the possibility of the
regularization which is UV sensitive. In the papers
\cite{Nomura,Weiner} it was claimed that the quadratically
divergent result obtained in \cite{Nilles,Kim} is because the
sharp momentum cutoff is not preserving the underlying symmetry,
`local supersymmetry' \cite{BHN2}. However, it is hard to accept
the argument because we know that mass dependent regularization
shows the power law divergent properties of physical quantites
more clearly. Momentum cutoff itself also preserves supersymmetry
and shows an excellent cancellations of bosonic and fermionic
contribution in the supersymmetric and softly broken limit.
Locality of the extra dimension was also stressed, but it is clear
that for sufficient high momentum cutoff the violation of the
locality would be extremely tiny and the smearing effect would not
affect the result. Momentum cutoff of the theory is the scale
below which the locality is valid. Therefore, it is remained as a
puzzle and is still on the debate \cite{MSSS,CKR,KT,CPRT}.

In this paper we extend the observation given in \cite{Kim} and
obtain more precise result based on the regularization scheme
which keeps 5-D Lorentz invariance \cite{Weiner} manifestly. To
capture the UV physics above the compactification scale correctly,
it is necessary to maintain 5-D Lorentz invariance since we can
not distinguish the compactified extra dimension from other
noncompact 4 dimension at short distances. In other words short
distance physics does not discriminate KK modes from 4-D momentum,
and the correct regularization should also have this property. The
observation made in \cite{Kim} is not complete because the 4-D
momentum cutoff differs for each KK mode if we start from 5-D
momentum cutoff. Let $R$ be the compactification radius, $N$ be
the level of the KK modes such that KK modes heavier than $N/R$
are truncated, and $\Lambda$ be the cutoff of the 4-D momentum.
The precise relation in Eucildean space is $p_4^2 + p_5^2 \le
\Lambda^2$. Thus $N = [R\Lambda]$ is fixed with $[$ $]$ Gaussian
integer and the 4-D momentum of the $n$th KK mode has a cutoff
$p_{4_n}^2 \le \Lambda^2 - (\frac{n}{R})^2$ and $p_{4_N}$ should
be nearly zero. In this paper we calculate the effect of these
mode dependent cutoff motivated by 5-D Lorentz invariance and show
that the UV insensitive result is recovered. This is a distinct
feature of SS supersymmetry breaking mechanism.

\section{Scherk-Schwarz mechanism}

Scherk-Schwarz breaking of symmetry is realized by modifying the
periodic boundary condition along the compact extra dimension with
compactification radius $R$. For the circle compactification, \bea
\phi(x^{\mu}, y + 2\pi R) & = & e^{2\pi i a} \phi(x^{\mu}, y),
\eea where $0 \le a < 1 ({\rm{or}} -\frac{1}{2} < a \le
\frac{1}{2}$) is a parameter controlling the breaking of the
relevant symmetry. For maximal breaking of supersymmetry we give
$a = \frac{1}{2}$ for bosons and $a=0$ for its superpartners and
vice versa. The simplest realization of the effect is \bea
\phi(x^{\mu}, y) & = & \sum_{n=-\infty}^{\infty} e^{i(n+a)
\frac{y}{R}} \phi_n (x^{\mu}), \eea and the generalized
realization for orbifolding setup is in \cite{BFZ,BFZ2}. The
spectrum is modified as \bea m_{n B,F}^2 & = & (\frac{1}{R})^2
(n+a_{B,F})^2, \eea for bosons and fermions.

Scherk-Schwarz breaking of supersymmetry should be distinguished
from orbifolding breaking of it. {\footnote{The author thanks H.
P. Nilles for pointing out this.}} In order to get chiral theory
from higher dimensional theory, orbifolding procedure is essential
and $S^1/Z_2$ is assumed in which $Z_2$ is for $y \rightarrow -y$.
It is the common setup that can have chiral zero modes. $Z_2$
breaks $N=2$ to $N=1$ in 4-D viewpoint, and there is $N=1$
supersymmetry after the orbifolding and two fixed points. There
are two ways of breaking this $N=1$ further. One is the
orbifolding \cite{BHN} with ${Z'}_2$ and the other is SS breaking
of supersymmetry \cite{ADPQ,DPQ}. Apparently it looks different,
but as long as the spectrum is concerned, we can make the same
spectrum of orbifolding by taking $a=\frac{1}{2}$ in SS breaking.
For gauge theories, the equivalence of breaking by orbifold
boundary condition and by background gauge fields is discussed in
\cite{HMN}. Thus the model in \cite{BHN} is also considered as a
kind of SS breaking of supersymmetry.

Though our main concern is the one loop mass correction to the
Higgs from top and stop loop involving Yukawa couplings, we
consider more general expressions that hold for corrections
involving Yukawa couplings and gauge couplings. The expression is
given already for its analytic continuation to the Euclidean
space. We are interested in the generic properties of SS
supersymmetry breaking, and set $a_B = a$ and $a_F = 0$ for
simplicity. One loop mass correction to the scalar from bosons and
fermion loops is given by

\bea \label{loop1} m^2 & = & C g^2 A  \\ \label{loop2} A & = &
\sum_{n=-\infty}^{\infty} \int_0^{\infty} dp p^3 \left[
\frac{1}{p^2 + (n+a)^2} -\frac{1}{p^2+n^2} \right], \eea with $C$
a constant which is irrelevant for later discussion and $g$ is a
coupling constant (Yukawa or gauge). We use rescaled dimensionless
4-D momentum $p = R P$ of 4-D momentum $P$ and $\bar{\Lambda} = R
\Lambda$ of the 5-D momentum cutoff $\Lambda$ for notational
simplicity. Since each term is divergent, we have to regulate the
propagator. The first option is to use the Pauli-Villars
regulator. For massless fields, the regulator corresponds to \bea
\frac{1}{p^2} \rightarrow \frac{1}{p^2}
(\frac{\Lambda^2}{p^2+\Lambda^2}). \eea We can apply it to all of
the KK modes. \bea \frac{1}{p^2+m^2} \rightarrow
\frac{1}{p^2+m^2}\left(\frac{\Lambda^2}{p^2+m^2+\Lambda^2}\right)
= \left[ \frac{1}{p^2+m^2} - \frac{1}{p^2+m^2+\Lambda^2} \right].
\eea Sharp momentum cutoff can be understood as a limit of higher
PV regularization, $L \rightarrow \infty$ \bea \frac{1}{p^2+m^2}
\rightarrow
\frac{1}{p^2+m^2}\left(\frac{\Lambda^2}{p^2+m^2+\Lambda^2}\right)^L
= \left[ \frac{1}{p^2+m^2} - \sum_{l=1}^{L}
\frac{1}{(p^2+m^2+\Lambda^2)^l} \right]. \eea

Therefore, we get the correct sharp momentum cutoff in the limit
$L \rightarrow \infty$. This is slightly different from the
previous one \cite{Nilles,Kim}. In the previous calculation, four
dimensional momentum cutoff($\Lambda_4$) and KK momentum
truncation ($N$) have been taken independently \cite{Nilles}, and
the cutoff of the four momentum and the KK truncation scale have
been identified ($\Lambda_4 R = N$) \cite{Kim} using the physical
argument that they are constrained by the five dimensional cutoff.
Here we have an improved regularization which avoids unphysical
situation automatically. (anisotropic regularization
\cite{Kim}){\footnote{The regularization invented in this paper is
indebted to V. A. Rubakov. The author thanks him for his keen
suggestion.}

Though $L \rightarrow \infty$ corresponds to the sharp momentum
cutoff limit which preserves 5-D Lorentz invariance, the above
regulator is inadequate since it suppresses all the contributions
except extremely IR contribution. For example, near the cutoff
$P^2+m^2 \sim \Lambda$, the propagator is already
$(\frac{1}{2})^L$ suppressed and there remains rare correction.

Therefore we use different momentum dependent regularization that
keeps the contribution near the cutoff finite even in the sharp
cutoff limit. Let us start from the regulated propagator of 5-D theory

\bea \frac{1}{{\tilde{P}}^2} \rightarrow \frac{1}{{\tilde{P}}^2}
e^{-\frac{{\tilde{P}}^2}{\Lambda^2}}. \eea

Though it is nonlocal and there is no Hamiltonian formulation for
this modified propagator, it can be thought of as a perfectly good
regulated propagator. Furthermore, we can generalize it for
arbitrary integer $L > 0$ such that

\bea \frac{1}{{\tilde{P}}^2} \rightarrow \frac{1}{{\tilde{P}}^2}
e^{\{-(\frac{{\tilde{P}}^2}{\Lambda^2})^L\}}. \eea

Since here ${\tilde{P}}^2$ corresponds to the 5-D momentum square
and is decomposed into $P^2 + (\frac{n+a}{R})^2$. The KK mode
dependent regulator keeps 5-D Lorentz invariance manifestly. Now
near the cutoff, $p^2 + (n+a)^2 = {\bar{\Lambda}}^2$, the
propagator is just $1/e$ times the original one and is independent
of $L$. Thus we can take $L \rightarrow \infty$ limit while
keeping the near cutoff contributions finite. The smooth function
becomes a step function and the sharp momentum cutoff is realized.
In the discussion, $L \rightarrow \infty$ is not necessarily
important but just simplifies the proof. Actually all the
properties are maintained only if $L \ge \bar{\Lambda}
(=R\Lambda)$ since the step function can be an accurate
approximation for $L \ge \bar{\Lambda}$. The propagator is varied
rapidly near $\bar{\Lambda}$ just within the interval
${\bar{\Lambda}} - 1$ and $\bar{\Lambda} + 1$ for $L \ge
\bar{\Lambda}$. Having this keeping in mind, we use $L \rightarrow
\infty$ limit in the following discussion. The sum is truncated at
$N = [\bar{\Lambda}]$ and the 4-D momentum cutoff becomes \bea C_n
& = & \sqrt{({\bar{\Lambda}}^2 - (n+a)^2)}. \eea

The one loop contribution in eq. (\ref{loop1}),(\ref{loop2}) is

\bea A & = & A(a) - A(0), \\
 A(a) & = & \sum_{n=-N}^{N} \int_0^{C_n} dp
\frac{p^3}{p^2+(n+a)^2} \\ & = & \sum_{n=-N}^{N} \left[
\frac{1}{2} C_n^2 - \frac{1}{2} (n+a)^2 \log (\frac{C_n^2 +
(n+a)^2}{(n+a)^2}) \right] \\ & = & \sum_{n=-N}^{N} \left[
\frac{1}{2} \left( {\bar{\Lambda}}^2 - (n+a)^2 \right) -
\frac{1}{2} (n+a)^2 \log \frac{{\bar{\Lambda}}^2}{(n+a)^2} \right]
.\eea

Direct summation over $N$ with $\bar{\Lambda} = N+\frac{1}{2}$ and
Stirling's formula give us a simple expression in which $N^3, N^2,
N$ terms are cancelled.

\bea A(a)  = \sum_{n=-N}^{N}
&& \left[ \frac{1}{2} (N+\frac{1}{2})^2
-\frac{1}{2} (n+a)^2 \right. \nn \\
&& \left. - \frac{1}{2} (n+a)^2
\log \frac{(N+\frac{1}{2})^2}{n^2(1+\frac{a}{n})^2} \right] \eea

\bea A & = & A(a) - A(0) = -\frac{5}{12} \zeta(2) a^2
+ {\cal O}(\frac{a^2}{N^2}, a^4) \nn \\
& = & -\frac{5\pi^2}{72} a^2
+ {\cal O}(\frac{a^2}{N^2}, a^4) \eea

From the expression, cubic divergence appears both in the bosonic
and fermionic loops and cancel with each other. The final result
does not depend on any positive powers of $N$ (or $\bar{\Lambda}$,
the UV cutoff) and $\log N$. This is a special property of
Scherk-Schwarz supersymmetry breaking. Here the cancellation of
quadratic divergence in the bosonic and the fermionic sector
respectively is easy to understand from the property of SS
supersymmetry breaking. If we expand $A(a)$ in terms of $a$, we
get \bea A(a) & = & \sum_{i=0}^{\infty} A_i a^i =  A_0 ({\cal O}
(N^3)) + A_2 ({\cal O} (N)) a^2 + {\cal O}(a^4). \nn \eea
Kaluza-Klein towers have a $Z_2$ symmetry {\footnote{Theories with
$a$ and $-a$ have the same spectrum.}} which changes $a
\rightarrow -a$ and this prevents terms with odd powers of $a$
($A_1$, $A_3$, $\cdots$). {\footnote{ We assume $-\frac{1}{2} < a
< \frac{1}{2}$ in order to avoid unnecessary subtlety related to
the mode located just at the boundary. For $a = \frac{1}{2}$, we
can think it as the limit starting from $b = a - \epsilon$ with a
tiny $\epsilon > 0$. This avoids the potential danger of sharp
momentum cutoff and makes the sharp momentum cutoff good regulator
preserving supersymmetry. Here it is easily maintained just by
keeping the numbers of bosons and fermions equal. However this
caution does not affect the conclusion because the contribution of
these edge KK modes is extremely suppressed due to the constraint
on 4-D momentum $p^2 \le \Lambda^2 - N^2 \sim 0$.}} Therefore net
contribution of boson and fermion is \bea A & = & A_2 ({\cal O}
(N)) a^2 + {\cal O}(a^4). \nn \eea Apparently there are
contributions which are linearly divergent ${\cal O}(N)$ (${\cal
O}(\Lambda)$) and logarithmically divergent ${\cal O}(\log N)$
(${\cal O}(\log \Lambda)$). It is puzzling that these two
contributions are absent and the final result does not have UV
sensitivity.

There are more convenient approach for general proof. When $N$ is
large enough ($\bar{\Lambda}(=R\Lambda) \gg 1$), we can
approximate the sum over finite modes by a definite integral over
finite range.

\bea \sum_{n=-N}^{N} f(n+a) & = & \int_{-N}^N dx ~f(x+a) +
\frac{1}{2} \left[f(N+a)+f(-N+a)\right] \nn \\ &&+
\sum_{m=1}^{L-1} \frac{1}{(2m)!} B_{2m} \left\{ f^{(2m-1)}(N+a) -
f^{(2m-1)}(-N+a) \right\} \nn \\ &&+ \frac{1}{(2L)!} B_{2L}
\sum_{n=-(N-1)}^{N-1} f^{(2L)} (a + n + \theta) , \eea for some $0
< \theta <1$ and $f^{(m)} = \frac{d^m f}{dx^m}$.
This is the Euler-Maclaurin formula for $f(x)$ whose
first $2L$ derivatives are continuous on the interval
$(-N+a,N+a)$.

The general proof for the softness of SS supersymmetry breaking is
following. If there are quantities with the following structure
\bea A & = & \sum_{n=-N}^{N} \left[ f(n+a) - f(n) \right], \eea
then $A$ does not depend on $N$ if $f(n)$ satisfies a few
conditions. To be more specific in the 5-D case, \bea f(x) & = &
\frac{1}{2} ( \bar{\Lambda}^2 - x^2) - \frac{1}{2} x^2 \log
(\frac{\bar{\Lambda}^2}{x^2} ). \eea Since $f(x)$ is even
function, we have \bea f(-x) & = & f(x),  \\ f^{(2n-1)} (0) & = &
0. \eea Using the Euler-Maclaurin formula and the special
properties of $f(x)$, we obtain after the Taylor series expansion
around N \bea A = \sum_{m=1}^{\infty} &&\left[
\frac{f^{(2m-1)}(N)}{(2m)!} a^{2m} + \frac{2 f^{(2m)} (N)}{(2m)!}
a^{2m} + \frac{2 f^{(2m)} (0)}{(2m)!} a^{2m} \right. \nn \\
&&\left. + \sum_{k=1}^{L-1} \frac{2 B_{2k} f^{(2k-1+2m)}(N)}{(2k)!
(2m)!} a^{2m} \right], \eea where $B_n$ is the Bernoulli number.

Let us look at the potentially UV dependent terms. Since $f(x)$ is
quadratic in $x$, to investigate the behavior of $f^{(1)} (x)$ and
$f^{(2)}(x)$ is enough for our purpose of checking UV dependence.
($f^{(1)}(x)$ can be ${\cal O} (N)$ or ${\cal O}(N \log N)$ and
$f^{(2)}(x)$ can be ${\cal O}(\log N)$. All $f^{(n)}(x)$ with
$n>2$ are ${\cal O}(1)$ or less.) The first term contains $f^{(1)}
(N) a^2$ term but $f^{(1)} (N) = {\cal O} (1)$ because there is
little room for 4-D momentum space of higher KK modes near the
cutoff. The second and the third term contain $f^{(2)} (N) a^2$
and $f^{(2)} (0) a^2$. Therefore, to check $f^{(2)} (N)$ and
$f^{(2)} (0)$ is enough to insure that there are no UV dependence
in the mass correction. It turns out that $f^{(2)} (N) = {\cal
O}(1)$ and $f^{(2)} (N) = {\cal O}(1)$ and there is no $\log N$
(or $\log \Lambda$) dependence.

As is shown above, the contribution of KK modes from SS
supersymmetry breaking shows extremely soft nature. 5-D Lorentz
invariance naturally suppresses potentially dangerous
contributions of heavy KK modes (near the cutoff) by reducing the
available 4-D momentum space for those modes. It is natural to ask
whether this is special in 5-D or generic in all dimensions. In
7-D model, $f(x) = {\cal O}(x^4)$ and $f^{(1)} (N)$, $f^{(3)}
(N)$, $f^{(2)} (N)$, $f^{(2)} (0)$, $f^{(4)} (N)$ and $f^{(4)}
(0)$ are shown to be ${\cal O}(1)$. These are enough to show that
there are no UV dependence for the mass correction of the Higgs.
Though we have not addressed the general proof, it is very
plausible that in all dimensions, the SS supersymmetry breaking is
extremely soft.

The properties of SS breaking \cite{MP,KW} looks plausible from
Wilson line interpretation of the gauge symmetry breaking
\cite{MSSS,HMN}. In that case nontrivial boundary condition can be
identified to the configuration with constant background gauge
field with periodic boundary condition. The constant gauge field
can be gauged away locally, and it becomes a physical one only for
the compact space. UV sensitivity is determined from the local
physics, UV insensitivive result for SS symmetry breaking looks
very natural since we can locally gauge it away.

\section{Conclusion}

Starting from the regulator preserving 5-D Lorentz invariance, we
calculated one loop correction to the scalar mass with SS
supersymmetry breaking spectrum in the extra dimensional model. On
the contrary to the anticipation from the naive physical isotropic
regularization \cite{Kim} or truncation with momentum cutoff
\cite{Nilles}, the result is UV insensitive due to the special
property of SS supersymmetry breaking. Soft breaking of
supersymmetry in the bulk reduces the degree of divergence by two
powers with the aid of boson-fermion cancellation, but
Scherk-Schwarz breaking gets rid of all the divergences, i.e., UV
sensitivities. The result depends only on the compactification
radius and the SS supersymmetry breaking parameter. Sharp momentum
cutoff is a good regularization consistent with supersymmetry as
long as 5-D Lorentz invariance is preserved. The UV insensitive
answer obtained by (apparently) unphysical KK regularization is
confirmed to be right in the momentum cutoff regularization which
is more physical and does not contain any disputable procedure.

While this paper was almost finished, three papers
\cite{GNS,GNN,N} appeared. The shift symmetry in those papers is
naturally realized in the regulator used in this paper before
sending $L \rightarrow \infty$. The shift symmetry is the discrete
version of the boosting symmetry for the compact extra dimension,
and 5-D Lorentz symmetry contains it.

\section*{Acknowledgments}

The author thanks K. Choi, Y. Kiem, J. E. Kim, P. Ko and H. P.
Nilles for reading a draft and for useful comments and especially
V. A. Rubakov for suggesting me to refine the anticipation. The
author also thanks the participants in SUSY'01 for discussions.

\end{document}